\documentclass[10pt,aps,twocolumn,showpacs,preprintnumbers,amsmath,amssymb,prl,
superscriptaddress,floatfix]{revtex4-1}

\usepackage{graphicx}

\begin{document}

\title{Current Correlations in Quantum Spin Hall Insulators}

\author{Thomas~L.~Schmidt}\affiliation{Department of Physics, Yale University, 217 Prospect Street, New Haven, CT 06520, USA}

\date{\today}

\newcommand{\expct}[2]{\left\langle #1 \right\rangle_{#2}}
\newcommand{\expcts}[2]{\langle #1 \rangle_{#2}}
\newcommand{\cumu}[2]{\langle\!\langle #1 \rangle\!\rangle_{#2}}
\newcommand{\hc}{\text{h.c.}}
\newcommand{\T}{\mathcal{T}}
\newcommand{\C}{\mathcal{C}}
\newcommand{\TC}{T_{\C}\, }

\begin{abstract}
We consider a four-terminal setup of a two-dimensional topological insulator (quantum spin Hall insulator) with local tunneling between the upper and lower edges. The edge modes are modeled as helical Luttinger liquids and the electron-electron interactions are taken into account exactly. Using perturbation theory in the tunneling, we derive the cumulant generating function for the inter-edge current. We show that different possible transport channels give rise to different signatures in the current noise and current cross-correlations, which could be exploited in experiments to elucidate the interplay between electron-electron interactions and the helical nature of the edge states.
\end{abstract}

\pacs{71.10.Pm, 72.15.Nj, 85.75.-d, 73.43.-f}

\maketitle

Topological insulators are characterized by having a nonzero band gap in the bulk while gapless edge modes exist on the surfaces \cite{hasan10,qi10}. The edge modes are protected by time-reversal symmetry and are robust against perturbations such as impurity scattering or electron-electron interactions. Two-dimensional topological insulators have been realized using HgTe quantum wells \cite{koenig07} and have been shown to display the behavior predicted for quantum spin Hall insulators \cite{roth09}: a strong spin-orbit coupling locks the spatial motion of the electron to its spin and thus entails the formation of \emph{helical} edge modes, where electrons with opposite spin orientations on any given edge propagate into opposite directions.

In the presence of electron-electron interactions, the edge modes of two-dimensional topological insulators can be modeled as helical Luttinger liquids (LLs) \cite{xu06,wu06,teo09}. This extension of the conventional LL theory provides a description of the low-energy degrees of freedom and makes an exact treatment of the interaction possible. In particular, it allows to analyze the nonequilibrium transport properties of helical edge modes \cite{liu11}. Some of these transport properties have already been investigated in experiments \cite{koenig07,roth09}.

To date, most attention has been devoted to the investigation of the average current through edge modes as a function of an applied bias voltage \cite{liu11,dolcini11}. However, it has been known for a long time that a measurement of the current noise yields additional information, e.g., about the nature of the charge carriers, which is not accessible from current measurements \cite{schottky18}. In this paper, we shall investigate the transport properties of edge states of two-dimensional topological insulators using full counting statistics (FCS), which yields insights into the average current, the current noise, and arbitrary higher-order correlation functions of the current.

\begin{figure}[ht]
  \centering
  \includegraphics[width = 0.48 \textwidth]{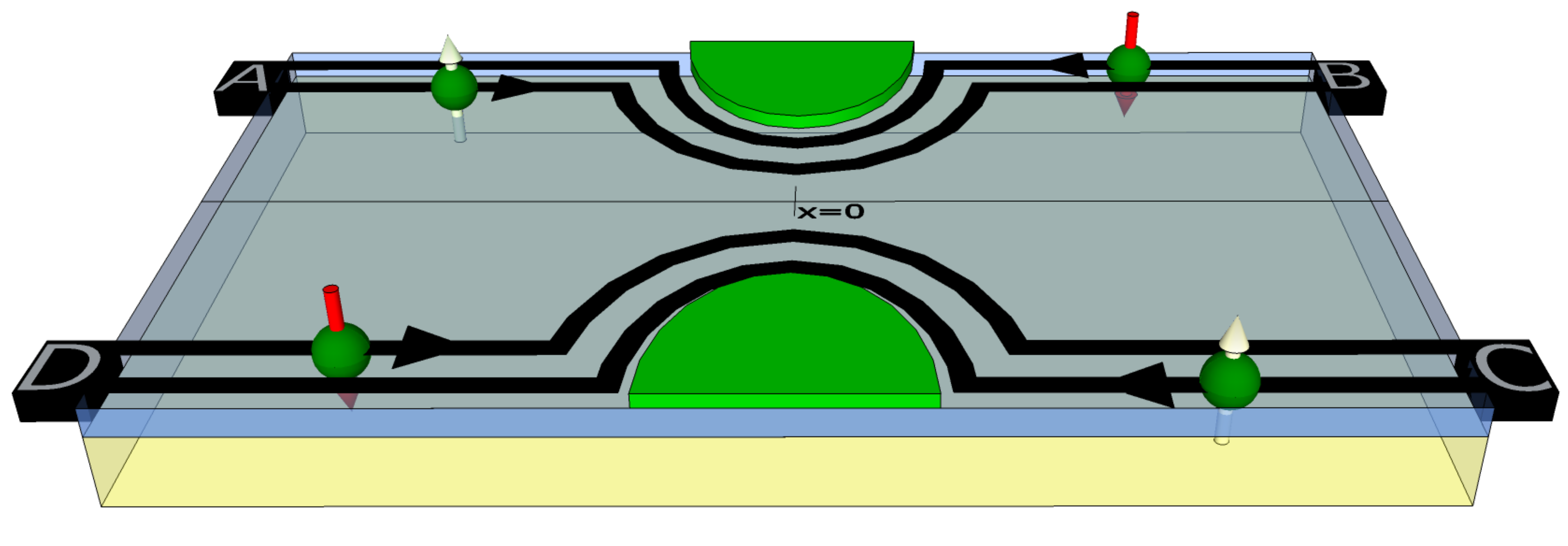}
  \caption{(Color online) The setup under investigation consists of a two-dimensional topological insulator with two helical edge modes. The four corners are connected to electron reservoirs held at bias voltages $\mu_{A,B,C,D}$. Using top gates, the upper and lower channels can be coupled locally, leading to inter-edge tunneling at $x=0$.}
  \label{fig:HallBar}
\end{figure}

We consider the two-dimensional topological insulator depicted in Fig.~\ref{fig:HallBar}. The system is insulating in the bulk, but the upper and lower edges of the sample support counter-propagating gapless helical modes. The upper edge contains a right-moving spin-up mode (denoted by the fermionic operator $\psi_{R\uparrow}$) and a left-moving spin-down mode ($\psi_{L\downarrow})$, while the lower edge hosts the modes $\psi_{R\downarrow}$ and $\psi_{L\uparrow}$. The two modes on each edge are coupled by electron-electron interactions, but in a time-reversal invariant system, these interactions cannot lead to backscattering. The helical LL theory is based on the bosonization of the fermionic operators (for $\alpha = R,L$ and $\sigma = \uparrow,\downarrow$) \cite{giamarchi03}
\begin{align}\label{boson}
 \psi_{\alpha\sigma}(x) = \frac{U_{\alpha\sigma}}{\sqrt{2\pi a}} e^{-i \varphi_{\alpha\sigma}(x)},
\end{align}
where $U_{\alpha\sigma}$ denote Klein factors which ensure the correct anticommutation relations between the fermionic fields \cite{haldane81} and $a$ is a short-distance cutoff. The helical LL theory can be formulated in different bosonic bases: the helical edge basis \cite{strom09} is the canonical choice and respects the separation between the upper and lower edges by using the fields $\phi_1 \propto \varphi_{R\uparrow} + \varphi_{L\downarrow}$ and $\phi_2 \propto \varphi_{L\uparrow} + \varphi_{R\downarrow}$. However, once tunneling between the two edges is considered, the spin-charge basis \cite{hou09} becomes more convenient. The latter mixes states on the two edges and describes the system in terms of spin and charge degrees of freedom by introducing the spin modes ($\phi_s$, $\theta_s$) and charge modes ($\phi_c$, $\theta_c$). For $\alpha = R,L = +,-$ and $\sigma = \uparrow,\downarrow = +,-$ these are defined by \cite{hou09}
\begin{align}
 \varphi_{\alpha\sigma} = \frac{1}{\sqrt{2}} \left( \alpha \phi_c - \theta_c + \alpha \sigma \phi_s - \sigma \theta_s \right).
\end{align}
The fields $\phi_\nu$ and $\theta_\nu$ ($\nu = c,s$) are canonically conjugate, $\left[ \phi_\nu(x), \partial_y \theta_\mu(y) \right] = i \pi \delta_{\mu\nu} \delta(x-y)$. In terms of these fields, the edge state Hamiltonian becomes a sum of independent spin and charge terms. The effect of electron-electron interactions manifests itself in the Luttinger parameters $K_c$ and $K_s$ for the charge and spin sectors. It was found that in a helical LL, these parameters are related by \cite{hou09,teo09}
\begin{align}
 K_c = \frac{1}{K_s} =: g,
\end{align}
where $g=1$ corresponds to noninteracting fermions and $0 < g < 1$ for repulsive interactions.

We assume that the four corners of the sample are connected to electron reservoirs held at chemical potentials $\mu_{A,B,C,D}$ as shown in Fig.~\ref{fig:HallBar}. In order to incorporate these voltage sources into our theory, we use the $g(x)$-model \cite{ponomarenko95,safi95,maslov95}. In this approach, the interacting region of the 1D wires extends from $x=-L/2$ to $x = L/2$ where $L$ is the sample length. The electron reservoirs, on the other hand, are modeled as noninteracting wires located at $|x| > L/2$. This approach has been used successfully to calculate current and noise properties of LLs \cite{dolcini05}. In the spin-charge basis, the Hamiltonian describing the interacting edge states thus becomes $H_0 = H_c + H_s$, where (using $e = \hbar = 1$ throughout this paper)
\begin{align}
 H_c &= \frac{v_F}{2\pi} \int_{-\infty}^\infty dx \left[ (\partial_x \theta_c)^2 + \frac{1}{g^2(x)} (\partial_x \phi_c)^2 \right] \\
 H_s &= \frac{v_F}{2\pi} \int_{-\infty}^\infty dx \left[ \frac{1}{g^2(x)} (\partial_x \theta_s)^2 + (\partial_x \phi_s)^2 \right]
\end{align}
where $g(x) = g$ for $|x| \leq L/2$ and $g(x) = 1$ for $|x| > L/2$. The voltage sources shift the chemical potentials in the noninteracting parts of the wire and affect both spin orientations in the same way. For instance, the voltage $\mu_A$ leads to a term $\mu_A \int_{-\infty}^{-L/2} dx [\rho_{R\uparrow}(x) + \rho_{L\downarrow}(x)]$. Analogous expressions proportional to $\mu_{B,C,D}$ appear at the three remaining corners. Using bosonization, the voltage sources can be described by the Hamiltonian
\begin{align}
 H_V = \frac{1}{\sqrt{2} \pi} \int dx \left[ E_c(x) \phi_c(x) + E_s(x) \theta_s(x) \right].
\end{align}
Note that the applied voltages couple to the charge field $\phi_c$ but to the \emph{conjugate} spin field $\theta_s$. This reflects the nonequilibrium spin-charge duality found in Ref.~\cite{liu11}. The voltage terms in the charge and spin sector are given by
\begin{align}
 E_{c,s}(x) &= (\mu_C \pm \mu_B) \delta(x - L/2) - (\mu_D \pm \mu_A) \delta(x + L/2).
\end{align}
Finally, we assume that near the point $x=0$, the setup is constricted in such a way that the helical modes on both edges spatially approach each other. In this case, local tunneling becomes possible. The possible tunneling terms in the presence of interactions in the wire were identified in Ref.~\cite{teo09}. First, spin-conserving single-particle tunneling leads to
\begin{align}
 T_e &= \gamma_e \sum_{\sigma} \psi^\dag_{R\sigma} \psi_{L\sigma} + \hc
\end{align}
where $\psi_{\alpha\sigma} \equiv \psi_{\alpha\sigma}(x=0)$. In addition, there may be tunneling of either charged or spinful particle pairs, which gives rise to the following terms
\begin{align}\label{Tc}
 T_c &=  \gamma_c \psi^\dag_{L\uparrow} \psi^\dag_{L\downarrow} \psi_{R\uparrow} \psi_{R\downarrow} + \hc, \\
 T_s &=  \gamma_s \psi^\dag_{R\downarrow} \psi^\dag_{L\uparrow} \psi_{R\uparrow} \psi_{L\downarrow} + \hc \label{Ts}
\end{align}
An analysis of the scaling dimensions of these three terms reveals that $T_e$ is the dominant contribution for weak interactions ($g \approx 1$). Towards stronger interactions, however, their magnitudes may becomes comparable and it is generally no longer justified to neglect pair tunneling \cite{teo09}. For $1/2 < g < 2$, all terms remain irrelevant in the renormalization group (RG) sense such that perturbation theory in $T_{e,c,s}$ is applicable. Note that using a mapping between the limiting cases of weak and strong inter-edge tunneling, our results can also be applied to the case where the sample is almost pinched off and only residual tunneling connects the left and right sides \cite{teo09}.

Full counting statistics (FCS) is a convenient tool to extract information about the nonequilibrium transport properties of this system. The objective of FCS is the calculation of the cumulant generating function (CGF) $\ln \chi(\lambda)$, which in its simplest form yields the average current, the zero-frequency noise and other current correlation functions. We are interested in calculating the statistics of the tunnel current, so the CGF is defined as
\begin{align}\label{lnchi_BS_def}
 \ln \chi(\lambda) = \ln \expct{ \exp\left[i \sum_{\alpha\sigma} \lambda_{\alpha\sigma} \delta N_{\alpha\sigma} \right]}{},
\end{align}
where $\delta N_{\alpha\sigma}$ denote the number of particles which have been scattered out of the edge mode  $\psi_{\alpha\sigma}$ during the measurement time $\T$. For large $\T$, the CGF becomes proportional to $\T$. Derivatives of Eq.~(\ref{lnchi_BS_def}) with respect to the counting fields $\lambda_{\alpha\sigma}$ then allow a calculation of the average tunnel currents $\expcts{\delta I_{\alpha\sigma}}{} = \expct{\delta N_{\alpha\sigma}}{}/\T$ and of arbitrary correlation functions, e.g., $\cumu{\delta N_{\alpha\sigma} \delta N_{\alpha'\sigma'}}{}$ which is proportional to the zero-frequency current noise.

A direct calculation of the CGF using the definition (\ref{lnchi_BS_def}) is difficult. It helps to take advantage of its representation as a time-ordered expectation value on the Keldysh contour \cite{levitov04},
\begin{align}\label{lnchi_BS_K}
 \ln \chi(\lambda) &= \ln \expct{ \TC \exp \left[ - i \int_\C ds T^{\lambda}(s) \right] }{V}.
\end{align}
Here, $\TC$ denotes the time-ordering operator on the Keldysh contour $\C$, which consists of two branches: $\C_-$ reaches from $-\infty$ to $+\infty$ whereas $\C_+$ leads back to $-\infty$ \cite{landau82}. The subscript $V$ in the expectation value indicates that it is taken with respect to the ground state of the Hamiltonian $H_0 + H_V$. In order to construct the tunneling operator $T^\lambda$, all fermionic operators have to be furnished with counting fields,
\begin{align}\label{psi_lambda}
 \psi_{\alpha\sigma}(x) \to e^{-i \lambda_{\alpha\sigma}/2} \psi_{\alpha\sigma}(x).
\end{align}
The counting fields have a nontrivial time-dependence: $\lambda_{\alpha\sigma}(t) = \pm \lambda_{\alpha\sigma}$ for $t \in \C_\pm$ \cite{levitov04}. Then, $T^\lambda$ can be constructed by using the substitution (\ref{psi_lambda}) in the definition of the tunneling Hamiltonian $T_e + T_c + T_s$, and bosonizing it using Eq.~(\ref{boson}).

Unfortunately, an exact calculation of the expectation value in Eq.~(\ref{lnchi_BS_K}) in nonequilibrium and in the presence of electron-electron interactions is not possible. Therefore, we consider the tunneling as a small perturbation. Then, we can use a linked-cluster expansion of Eq.~(\ref{lnchi_BS_K}) and truncate the series after the second (leading) order in $\gamma_{e,c,s}$. Up to this order, the total CGF remains a sum of three contributions due to the individual tunneling terms,
\begin{align}\label{chi_sum}
 \ln \chi = \ln \chi_e + \ln \chi_c + \ln \chi_s.
\end{align}
The single-particle term as well as the terms due to tunneling of charged pairs and spinful pairs are given by (using $\nu = c,s$)
\begin{align}\label{chi_parts}
 \ln \chi_e(\lambda)
&= \frac{\gamma_e^2 \T}{(2\pi a)^2} \int_{-\infty}^\infty ds \sum_{m = \pm} \sum_{\sigma=\uparrow,\downarrow = \pm}
 \left( e^{i m \sigma \lambda_{\sigma} } - 1\right) \notag \\
 &\times C_{c,m}(ms) C_{s,m\sigma}(ms), \\
 \ln \chi_\nu(\lambda)
&=
  \frac{\gamma_\nu^2 \T}{(2\pi a)^4} \int_{-\infty}^\infty ds \sum_{m = \pm}
      	( e^{i m \lambda_\nu} - 1) C_{\nu,2m}(ms), \notag
\end{align}
and depend on the following linear combinations of the counting fields, ($\sigma = \uparrow,\downarrow = +,-$)
\begin{align}
 \sigma \lambda_{\sigma} &= \lambda_{R\sigma} - \lambda_{L\sigma}, \notag \\
 \lambda_{c,s} &= \lambda_\uparrow \mp \lambda_\downarrow.
\end{align}
The CGF is determined by the bosonic correlation functions $C_{\nu,\zeta}(s)$. For $\nu = c,s$ and $\zeta \in \mathbb{Z}$, these are defined by
\begin{align}
 C_{\nu,\zeta}(t) &= \expct{e^{ \sqrt{2} i \zeta \phi_\nu(t)} e^{-\sqrt{2} i \zeta \phi_\nu(0)}  }{V}.
\end{align}
The calculation of these expectation values is a nontrivial task due to the position-dependence of the Luttinger parameter $g(x)$ and due to the presence of the voltage sources. However, the necessary techniques for ordinary and helical LLs have been developed and are explained in detail in Refs.~\cite{dolcini05,liu11}. The correlation functions are in principle known for arbitrary system length $L$ and even at nonzero temperatures. Finite-length effects in the tunnel current have been investigated in Ref.~\cite{liu11}. They have been found to lead to an oscillatory current-voltage characteristic which is observable only at very low bias voltages. Therefore, in order to keep the results compact, we shall focus on the case of zero temperature and infinite length.  In this case, the correlation functions become (for $\nu = c,s$)
 \begin{align}\label{Ccs_result}
 C_{\nu,\zeta}(t) &=  \exp\left\{ \frac{i \zeta \mu_\nu t}{2} \right\} \left[ \frac{ \left( \omega_c^{-1} + i t \right)^2 }{\omega_c^{-2} } \right]^{- K_\nu \zeta^2/2}
\end{align}
where $\omega_c = v_F/a$ is a high-energy cutoff and $K_c = 1/K_s = g$. The effective voltages $\mu_{c,s}$ affecting the charge and spin sectors are given by
\begin{align}
 \mu_{c,s} &= \frac{1}{2} (\mu_C \mp \mu_D \pm \mu_B - \mu_A ).
\end{align}
The time-integration in the CGF (\ref{chi_parts}) can be performed exactly and one finds the main result of this article: up to the second order in tunnel amplitudes, the CGF of the tunnel current consists of the terms,
\begin{align}\label{chi_final}
 \ln \chi_e(\lambda)
&= \frac{2^{-g - 1/g} \gamma_e^2 \T}{\pi a^2 \Gamma(g + 1/g)} \sum_{\sigma}  \left| \frac{(\mu_c + \sigma \mu_s)^{g + 1/g - 1}}{\omega_c^{g + 1/g}} \right| \notag \\
&\times \sum_{m = \pm}
 \left( e^{i m \sigma \lambda_{\sigma} } - 1\right)\theta[ m(\mu_c + \sigma \mu_s)],  \notag \\
 \ln \chi_\nu(\lambda)
&=
  \frac{\gamma_\nu^2 \T}{(2\pi)^3 a^4 \Gamma(4 K_\nu)}\left| \frac{\mu_\nu^{4 K_\nu - 1}}{\omega_c^{4K_\nu}} \right| \notag \\
&\times \sum_{m = \pm} ( e^{i m \lambda_\nu} - 1) \theta(m \mu_\nu)
 .
\end{align}
Hence, the CGF is a sum of three statistically independent Poissonian terms. The fact that all contributions are Poissonian is reasonable: by using perturbation theory in the tunneling, we assumed that inter-edge tunneling is a rare event. Next, we will discuss physical predictions for the current and current correlations which can be obtained from Eqs.~(\ref{chi_sum}) and (\ref{chi_final}).

The spin-resolved tunnel current can be calculated using $\expct{\delta I_\sigma}{} = \expct{\delta N_\sigma}{}/\T = (1/\T) \partial \ln \chi(\lambda) / \partial (i \lambda_\sigma) |_{\lambda = 0}$. Most notably, the three transport processes lead to different power-law current-voltage characteristics whose exponents are different combinations of the Luttinger parameter $g$. Power laws are a common feature of transport in interacting systems \cite{kane92}. The single-particle current has already been found in Ref.~\cite{liu11}. For small interactions $g \approx 1$, single-particle tunneling becomes linear in the applied voltage, and is thus dominant because pair-tunneling is cubic in $\mu_{c,s}$. Charged-pair tunneling, with an exponent $4g-1$, becomes more relevant towards stronger repulsive interactions ($1/2 < g < 1$). The exponent for spinful pair tunneling, on the other hand, is $4/g-1$, so its contribution is smaller than the other processes for repulsive interactions. All power laws are consistent with an RG analysis of the tunneling terms \cite{teo09}. The fact that all three contributions to the average current depend on different combinations of the applied voltages $\mu_{c,s}$ could allow an experimental distinction of the different transport processes.

The noise of the tunnel current is proportional to the second moment of the CGF, $S_\sigma = \cumu{\delta N_\sigma \delta N_\sigma}{} = (1/\T) \partial^2 \ln \chi(\lambda)/ \partial (i \lambda_\sigma)^2 |_{\lambda = 0}$. It is given by
\begin{align}
 S_\sigma
&= \frac{2^{-g - 1/g} \gamma_e^2}{\pi a^2 \Gamma(g + 1/g)} \sum_{\sigma}  \left| \frac{(\mu_c + \sigma \mu_s)^{g + 1/g - 1}}{\omega_c^{g + 1/g}} \right|  \\
&+
   \frac{\gamma_c^2}{(2\pi)^3 a^4 \Gamma(4 g)}\left| \frac{\mu_c^{4 g - 1}}{\omega_c^{4g}} \right|
+
  \frac{\gamma_s^2}{(2\pi)^3 a^4 \Gamma(4/g)}\left| \frac{\mu_s^{4/g - 1}}{\omega_c^{4/g}} \right|. \notag
\end{align}
Similar to the average current, the contributions due to single-particle tunneling, charged pair tunneling and spinful pair tunneling couple differently to the applied voltages $\mu_c$ and $\mu_s$. In the absence of scattering and at low temperatures, the transport in clean 1D LLs does not produce zero-frequency shot noise \cite{dolcini05}. Therefore, the tunnel current noise can directly be measured as the noise in either of the four contacts.

\begin{figure}[t]
  \centering
  \includegraphics[width = 0.48 \textwidth]{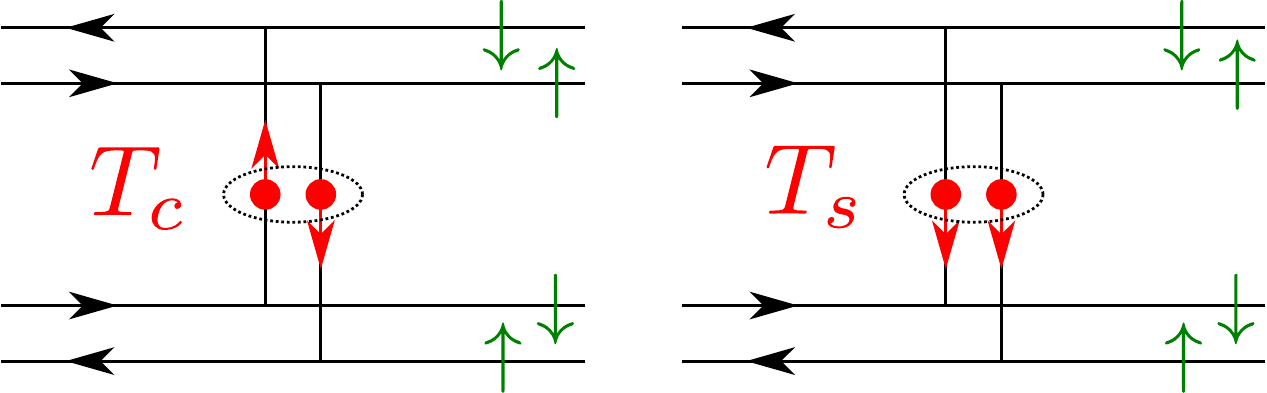}
  \caption{(Color online) Graphical depiction of charged-pair tunneling ($T_c$) and spinful-pair tunneling ($T_s)$ according to Eqs.~(\ref{Tc})-(\ref{Ts}). Due to the helicity, a single $T_c$ event turns, e.g., two right-movers into two left-movers and thus changes the charge transferred from left to right by two. $T_s$ affects the spin sector in an analogous way. Both processes give rise to cross-correlations between spin-up and spin-down tunnel currents $\cumu{\delta N_\uparrow \delta N_\downarrow}{}$. As $T_c$ transports two particles in opposite directions, whereas $T_s$ transports two particles in the same direction, the sign of the cross-correlations is different for both processes.}
  \label{fig:HBT}
\end{figure}

An even clearer distinction between the transport processes can be obtained by considering cross-correlations of the currents along different edges. These can be regarded as fermionic Hanbury Brown and Twiss (HBT) correlations and were shown previously to be significantly modified by interactions in multi-terminal setups \cite{schmidt07}. Such correlations have already been measured experimentally in quantum Hall bars \cite{henny99}. In the present setup, the interplay between interactions and the helical structure of the edge modes makes the result particularly interesting. Let us illustrate this with a simple bias configuration, $\mu_A = - 2 V$, whereas $\mu_{B,C,D} = 0$, such that $\mu_c = \mu_s = V$. We find
\begin{align}\label{HBT}
 S_\text{HBT}
&=
 \cumu{\delta N_{\uparrow} \delta N_{\downarrow}}{}
 \\
&=
 \frac{1}{(2\pi)^3 a^4} \left\{ \frac{\gamma_s^2}{\Gamma(4/g)}\left| \frac{V^{4/g - 1}}{\omega_c^{4/g}} \right|
-  \frac{\gamma_c^2}{ \Gamma(4 g)}\left| \frac{V^{4 g - 1}}{\omega_c^{4g}} \right| \right\}
  \notag
\end{align}
The single-particle tunneling does not show up in the HBT correlation because it does not couple spin-up and spin-down electrons. In contrast, particle-pair tunneling does yield HBT correlations. The sign of the correlations is positive for charged-pair transport and negative for spinful-pair transport. The physical reason for this behavior can be understood from the different forms of the tunneling terms, see Fig.~\ref{fig:HBT}.

Depending on the interaction strength, either charge transport or spin transport will be dominant \cite{teo09}. In the noninteracting limit ($g = 1$), $S_\text{HBT}$ is proportional to $|V|^3 (\gamma_s^2 - \gamma_c^2)$ and thus allows a direct comparison between the strengths of two pair-tunneling processes. For repulsive interactions, $1/2 < g < 1$, the different power laws associated with both processes will lead to a nonmonotonic $S_\text{HBT}(V)$. These features of the current cross-correlation are therefore direct evidence of pair tunneling and can be used to distinguish different transport regimes experimentally. Due to the helical structure of the edge modes, an experimental measurement of $S_\text{HBT}$ at this bias configuration does not require a spin-resolved current measurement. Instead, it can be determined by measuring the cross-correlation of the counter-propagating spin-up and spin-down currents on the lower edge, $S_\text{HBT} \propto \cumu{I_{L\uparrow} I_{R\downarrow}}{}$.

In realistic systems, spin-orbit coupling may also cause spin-flip tunneling processes \cite{teo09,liu11}, i.e., tunneling terms of the form $\psi^\dag_{R\uparrow} \psi_{R\downarrow}$. An analysis of such processes using FCS reveals that they do not cause correlations of HBT type. The physical reason is that spin-flip tunneling represents a single-particle process, whereas according to the definition (\ref{HBT}) correlated tunneling of two particles is required to produce HBT correlations. Therefore, the results for $S_\text{HBT}$ remain valid even in the presence of spin-flip tunneling. Moreover, all features should be visible for weak interactions, where the amplitudes for charged-pair and spinful-pair tunneling are comparable.

In conclusion, we have investigated transport in helical edge states of a two-dimensional topological insulator. For weak local tunneling between the upper and lower edges, we used perturbation theory to calculate the cumulant generating function for the tunnel current. This allowed us to analyze the current noise and current cross-correlations which emerge in this setup. Different transport processes produce strikingly different terms in the cross-correlations. This could be exploited to experimentally distinguish different transport regimes and investigate the interplay between electron-electron interactions and the helicity of the edge modes.

\acknowledgments The author wishes to thank A.~Komnik for valuable discussions and acknowledges support by the Swiss NSF.

\bibliography{paper}

\end{document}